\documentclass[onecolumn,showpacs,showkeys,prb,preprint]{revtex4}
\usepackage[sort&compress]{natbib}
\usepackage{graphicx}
\usepackage{dcolumn}
\usepackage{bm}
\usepackage{subfigure}       
\usepackage{hyperref}

\begin{document}

\preprint{Brazilian Journal of Physics/BJB}

\title{Energy dependence of a vortex line length near a zigzag of pinning centers}

\author{Mauro M. Doria}
 \email{mmd@if.ufrj.br}
\author{Antonio R. de C. Romaguera}%
 \email{ton@if.ufrj.br}
\affiliation{%
Instituto de F\'{\i}sica, Universidade
Federal do Rio de Janeiro\\
C.P. 68528, 21941-972, Rio de Janeiro RJ, Brazil
}%

\date{\today}

\begin{abstract}
A vortex line, shaped by a zigzag of pinning centers, is described
here through a three-dimensional unit cell containing two pinning
centers positioned symmetrically with respect to its center. The
unit cell is a cube of side $L=12\xi$, the pinning centers are
insulating spheres of radius $R$, taken within the range $0.2\xi$ to
$3.0\xi$, $\xi$ being the coherence length. We calculate the free
energy density of these systems in the framework of the
Ginzburg-Landau theory.
\end{abstract}

\maketitle

\section{Introduction}

Superconductors have many kinds of imperfections, that is, internal
regions where Cooper pairs either don't exist at all or exist purely
as a fluctuation effect. Thus the macroscopic wave function,
describing the collective state, vanishes abruptly or asymptotically
inside such regions. In presence of an external magnetic field
vortices arise inside the superconductor and are strongly attracted
to such imperfections, also called pinning centers. For this reason
pinning centers have been extensively studied in the past in many
ways, including artificially made ones, such as columnar
defects\cite{LGTBHP96}, antidots\cite{MBMRBTBJ98,YLBJW04} and micro
holes\cite{BP95}. They are interesting because they bring clear-cut
questions about the interaction between vortices and pinning
centers\cite{PM04,Bu93}, such as how local misalignment really
occurs inside the superconductor. Vortices lines should be aligned
to the applied field but the presence of strong attraction to a
pinning center can change this locally. As a result of competing
energetic demands new interesting phenomena can take place in vortex
Physics, such as the one considered here. A vortex line in the
absence of pinning centers is aligned along the magnetic induction
direction, hereafter called z-axis. The presence of a zigzag of
pinning centers forces the vortex line to bend and acquire this
shape, resulting into local misalignment, though it remains oriented
along the magnetic induction.

Pinning forces act on the vortex core whose radius is given by the
coherence length $\xi$. The interaction of pinning centers with
vortices has been studied using several approaches
\cite{KetSongbook}. From the point of view of the Ginzburg-Landau
theory, pinning may be caused by spatial fluctuations of the
critical temperature\cite{L70}, $T_c(\vec x)$, or of the mean
free-path\cite{B95} that changes the coefficient in front of the
gradient term, $\xi(\vec x)^2|({\vec \nabla} - {{2\pi
i}\over{\Phi_0}} {\vec A})\Delta|^2$. The interaction between a
vortex line and a pinning center has been considered by many authors
in the context of the Ginzburg-Landau theory\cite{B95,PF03}. The
number of vortex that can be trapped by a defect is an interesting
problem. In case of a columnar defect the saturation number has been
determined long ago by Mkrtchyan and Shmidt\cite{MS72}:
$n_s=\lambda/2\xi$, though this formula has to change near the upper
critical field\cite{DAS00}. Recently the saturation number has been
discussed for three dimensional cavities\cite{DZ02}.

The system under investigation here consists of a vortex line
trapped by a zigzag of very large pinning centers, namely, spherical
insulating cavities, with radius taken to vary from $0.2\xi$ to
$3.0\xi$. For these pinning centers the boundary-value problem has
to be taken into account, as usually treated since the de Gennes
boundary condition \cite{deGennesbook}must be satisfied at the
cavity surface. We chose to describe this system through a unit cell
with two cavities inside, that rotate freely around its
center, producing for each angle a distinct zigzag arrangement. In
this paper we analyze numerically the angular dependence of the
Helmholtz free energy $\mathcal{F}_c(\theta,R)$ and propose an
expression for it.

This paper is organized as follows. In section \ref{The model} we
present the model for a 3-D superconducting media with the cavities.
In section \ref{Theoretical approach}, we discuss out theoretical
approach. In sections \ref{Results} and \ref{A general case} we show
the results obtained through numerical simulations. In section
\ref{Conclusions} we summarize the main results of the work.

\section{The model}
\label{The model}

\begin{figure}[h]
\centering
\includegraphics[width=0.3\linewidth]{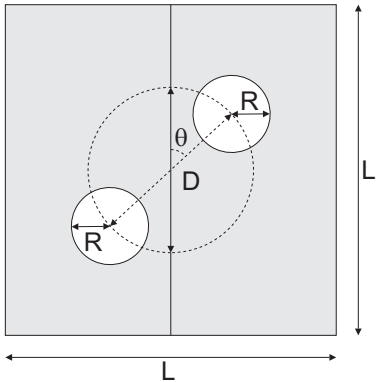}
\caption{The plane of rotation of the cavities inside the unit cell.
The distance $D$ between the two cavities is equal to $L/2$. The
superconductor fills the remain of the unit cell, shown here as a
gray region.} \label{figure1}
\end{figure}

The system studied here consists of a cubic
unit cell with size $L$ equal to $12\xi$, and two insulating
cavities of radius $R$, as specified by the figure \ref{figure1}.
The line segment joining the cavities makes an angle $\theta$
ranging from $0^\circ$ to $180^\circ$, taken here in increments of $3^\circ$.
Notice that because of symmetry $\theta$ varying from $0^\circ$ to $90^\circ$ is sufficient to
obtain all possible configurations.

%
%
%
%

Figure \ref{figure1} shows the plane of rotation (plane y) and
parameters associated to the unit cell. While $\theta$ and $R$ are
freely changed, $D$, the distance between the center of the
cavities, is fixed and equal to $L/2$. Thus the present model
features a distance between the two cavities independent of the
cavity radius.

Our numerical treatment demands a mesh grid to describe the unit cell.
The length of the cube side is $12\xi$, and we choose the number of mesh points
along a given direction, $P$, to be 19.
This choice implies that distance between two consecutive mesh points, $a$, be equal to $2\xi/3$.
The value of $a$ must be smaller than the coherence length $\xi$,
which is the minimum physical scale of the Ginzburg-Landau theory.
The number of grid points inside the cavity should be large enough
to describe it.
This number is obtained from the ratio between the two volumes,
of the cavity and of the cube.
Since the entire cube has $P^3$ points, the number of grid points inside the cavity is
$\frac{4}{3}\pi (\frac{R}{a})^3(1+\frac{a}{L})^3$.
For instance, a cavity of radius $R=1.0\xi$ has approximately 16 points, whereas
the $R=2.0\xi$ sphere has 8 times more points.
Obviously in the limit of a cavity with radius smaller thant the mesh distance, that is $R<a=2/3\xi$,
there will be just one point in the mesh describing the cavity.
In this case the order parameter does not
vanish inside the defect but just undergoes a drop on its value.

\section{Theoretical approach}
\label{Theoretical approach}

We start our considerations for the energy density functional of the
Ginzburg-Landau theory\cite{A57}, expressing it in units of the
critical field energy density\cite{DZ02}, $ H_c^2/4\pi $. The
periodicity of the problem requires a search of the free energy
minimum for a fixed integer, the number of vortices inside the unit
cell.

This integer also fixes the magnetic induction $\vec{B}(\vec{x})$,
which is the average of the local field taken over the unit cell
volume,
\begin{eqnarray}
\vec{B}=\frac{1}{v}\int_v \vec{h} d^3r.
\label{eq:localfield}
\end{eqnarray}
The magnetic induction is completely determined by the vorticity of
the system because the overall current circulation vanishes inside
the unit cell. The relationship between the vorticity of the system
$\vec{\nu}\phi_0$ and the magnetic induction $\vec{B}(\vec{x})$ in
reduced units is,
\begin{eqnarray}
\vec{B}(\vec{x})=2\pi\kappa\left(\frac{\xi}{L}\right)^2 \vec{\nu},
\label{eq:B}\end{eqnarray}
where $\vec{\nu}=n_x\hat{x}+n_y\hat{y}+n_z\hat{z}$ is the vorticity
in an arbitrary direction.
In the present paper we consider $\vec{\nu}=\hat{z}$ and present
some results concerning the $\vec{\nu}=2\hat{z}$ case.
The parameter $\kappa=\lambda/\xi$ is the
dimensionless Ginzburg-Landau parameter and $\lambda$ is the
penetration depth.
Here we consider the no magnetic shielding limit.
The field penetrates in the superconductor with no Meissner-Ochsenfeld effect.
In this regime $\vec{h}(\vec{x})=\nabla\times\vec{A}(\vec{x})= \vec{B}$.
This situation can be viewed as a large $\kappa$ limit.
In reduced units the free energy density is normalized by the
critical field density, $H_c^2/4\pi$, and the order parameter
density is dimensionless, varying between 0 and 1.
\begin{eqnarray}
\mathcal{F}_c = \int {{dv}\over{V}} \;\tau(\vec x) \left[ \xi^2
\left|({\vec \nabla} - {{2\pi i}\over{\Phi_0}} {\vec A})\Delta\right|^2 -
\left|\Delta\right|^2 \right] + {1 \over 2} \left|\Delta\right|^4, \label{eq:glth}
\end{eqnarray}
The function $\tau(\vec{x})$ is a step-like function used to specify
the cavities in this approach\cite{DZ02}. Explicitly we have
$\tau(\vec{x})=\tau_1(\vec{x})\tau_2(\vec{x})$ and
\begin{eqnarray}
\label{eq:tau} \tau_i({\vec x}) = 1 - \frac{2}{1 + e^{{(|\vec x -
\vec x_i|/R)}^{N}}},
\end{eqnarray}
where $\tau_i$ is equal to $0$ inside and $1$ outside the $i$th
cavity. The above explicit representation of the $\tau$ function is
necessary for computational reasons and for accuracy we take that
$N=8$. In the limit $N\to\infty$, the function $\tau$ tends to the
well-known Heaviside function, $\tau(\vec{x})=\Theta\left (
\frac{|\vec{x}-\vec{x}_1|}{R}-1\right)\Theta\left (
\frac{|\vec{x}-\vec{x}_2|}{R}-1\right)$.

Since we are in the no shielding limit, the vector potential
$\vec{A}(\vec{x})$ is determined from Eqs. \ref{eq:localfield},
\ref{eq:B} and the condition of magnetic flux quantization inside the unit cell.
The vector potential does not participate in the minimization process of
the free energy density that only takes into account the real and
imaginary parts of the order parameter. The free energy density
contains two terms. The first term is the condensation energy
density, $- \tau(\vec x)\left|\Delta\right|^2+ {1 \over 2}
\left|\Delta\right|^4$, which in case of no vortices
($\left|\Delta\right|^2=1$) and no cavity ($\tau=1$) has the value
-0.5. The presence of a cavity raises the energy since inside it the
density vanishes ($\left|\Delta\right|^2=0$). And the second term,
the kinetic energy density, $\tau(\vec x)\xi^2 \left|({\vec \nabla}
- {{2\pi i}\over{\Phi_0}} {\vec A})\Delta \right|^2$. Notice that
there is kinetic energy in case of no vortices but with a cavity. At
the insulating-superconducting interface $\tau$ changes from 1 to 0
and this causes a bending of the order parameter, which has some
kinetic energy cost.

The most significant advantage of the present method, is that the
free energy functional, Eq.~\ref{eq:glth}, contains the appropriate
boundaries conditions to the problem. This removes the necessity of
solving the theory in two independent regions and later applying the
Neumman boundary conditions. Besides the present method easily
applies to internal regions of any shape, not just spherical, and
finds its solution for the given normal-superconductor interface.

\section{Results}
\label{Results}

For a given pinning arrangement, which means fixed values of
$\theta$ and $R$, and the condition of one vortex in the unit cell,
we carry the minimization of the free energy. Here we present the
results of several independent simulations obtained for the pairs
($\theta,R$), ranging from $(0^\circ,0.2\xi)$ to
$(180^\circ,3.0\xi)$, in increments of $3^\circ$ for $\theta$,
and of $0.2\xi$ for $R$.
For each simulation we initialize the order parameter in a
random way and the minimization procedure is carried for each
temperature\footnote{The temperature here is the parameter
associated with the metropolis algorithm. Its meaning is not related
with the thermodynamical variable of the system.}. Then the
temperature is lowered until a convergence criteria is reached,
which means that the free energy has become stable. For each
simulation we implement at least 1200 Monte Carlo visits per mesh
point in a Metropolis algorithm.

\begin{figure*}[!t]
\centering
    \subfigure[]
    {
    \begin{minipage}[t]{0.46\textwidth}
    \centering
    \includegraphics[width=\linewidth]{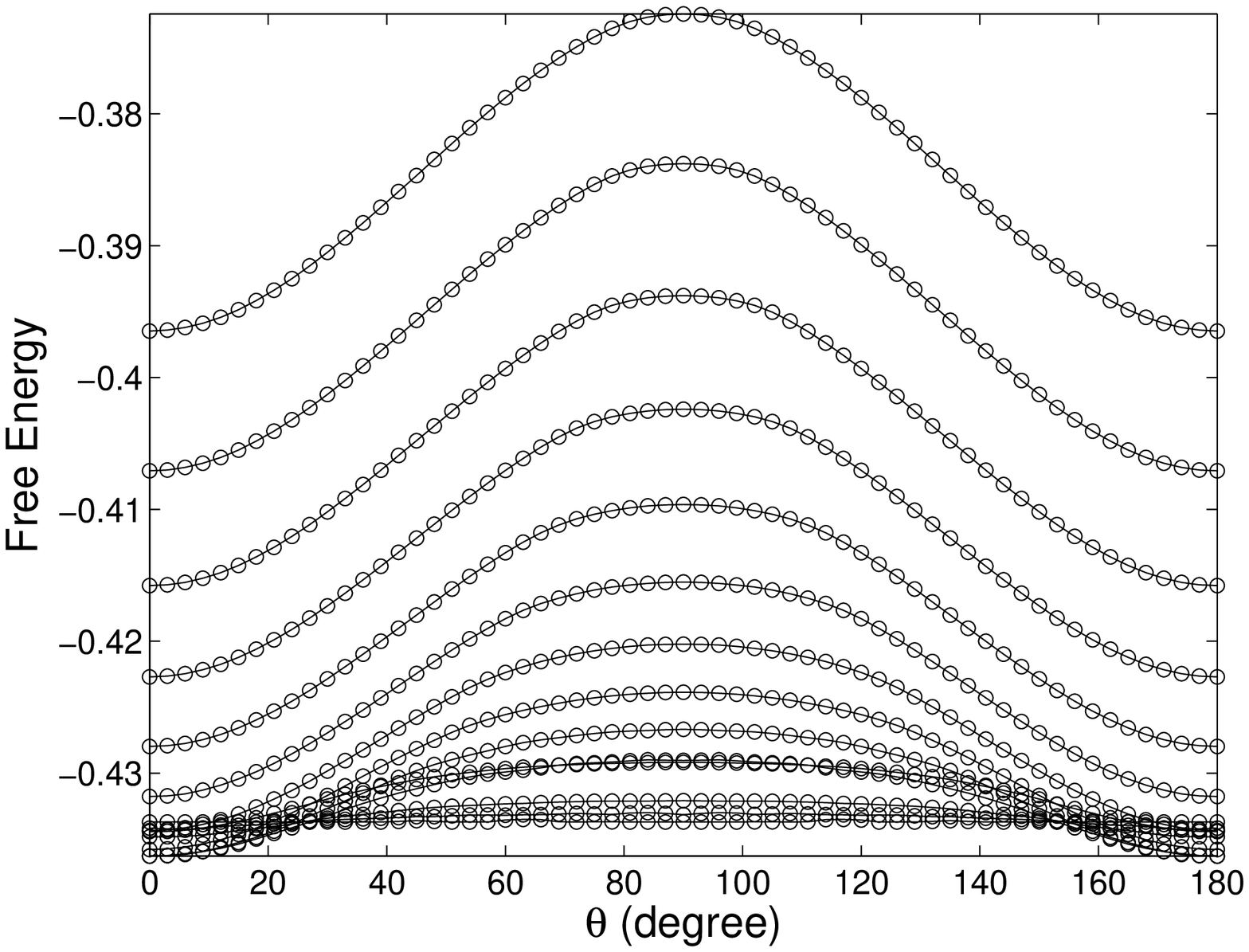}
    \label{figure2a}
    \end{minipage}
    }
    \subfigure[]
    {
    \begin{minipage}[t]{0.46\textwidth}
    \centering
    \includegraphics[width=\linewidth]{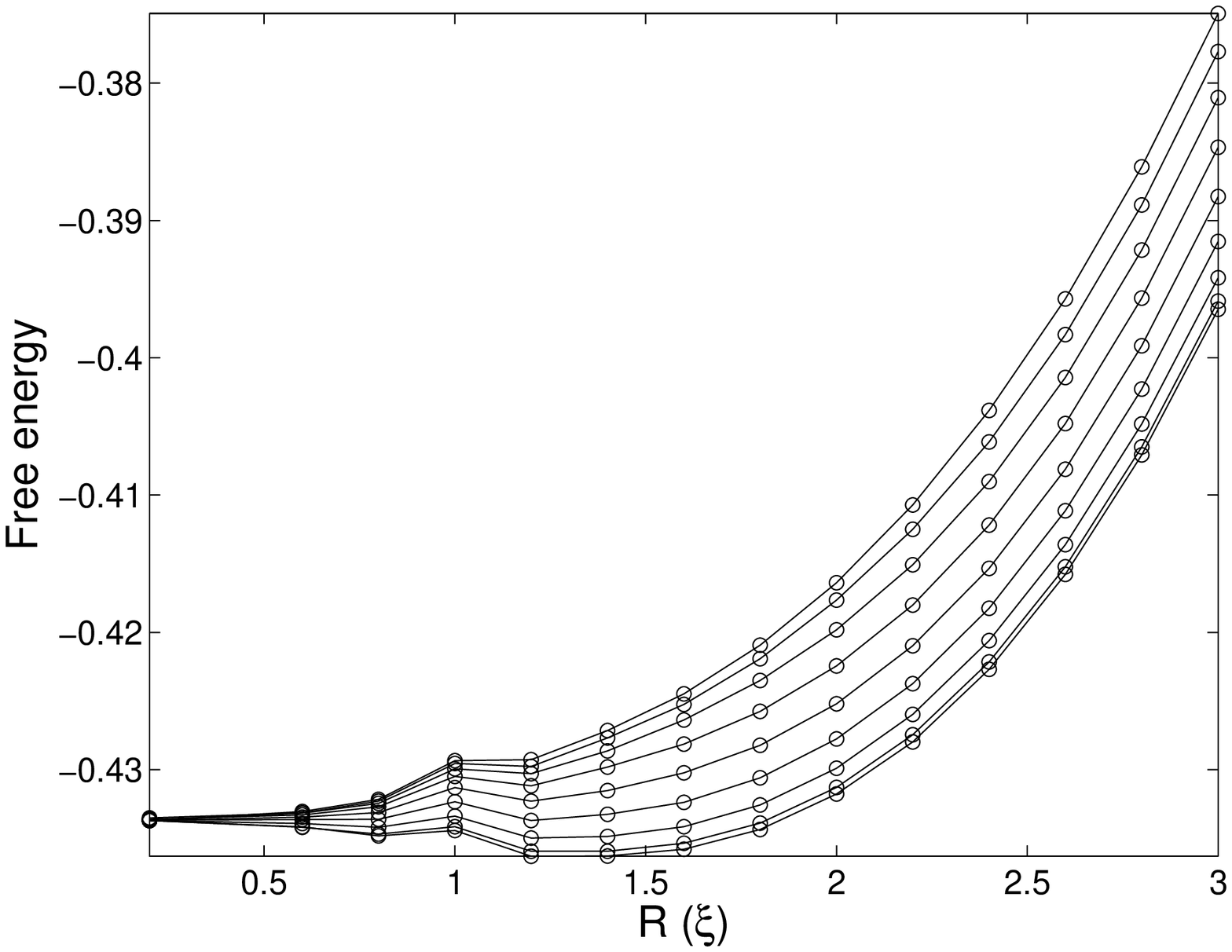}
    \label{figure2b}
    \end{minipage}
    }
    \caption
    {
    Dependence of the free energy, $\mathcal{F}_c$, with the pinning center radius $R$ and the
    angle $\theta$.
    \ref{figure2a}
      {Variation of
    $\mathcal{F}_c$ as a function of the angle $\theta$ in the range $0^\circ$ to $180^\circ$,
     data points obtained for increments of $3^\circ$. The radius $R$ varies from $0.2\xi$
    to $3.0\xi$ and for each increment of $0.2\xi$ results in a distinct curve,
    all plotted in ascendant order from bottom to top.}
    \\
    \ref{figure2b}
    {Variation of
    $\mathcal{F}_c$ as a function of the pinning center radius for a specific value of $\theta$.
    The radius varies from $0.2\xi$ to $3.0\xi$ with an increment of $0.2\xi$. Distinct curve correspond to different
    $\theta$, equal to $0^\circ$, $9^\circ$, $18^\circ$, $27^\circ$, $36^\circ$, $45^\circ$, $54^\circ$,
    $63^\circ$ and $72^\circ$ in the ascendant order form bottom to top.}
    }
    \label{figure2}
\end{figure*}

\begin{figure}[t]
\centering
\includegraphics[width=0.5\linewidth]{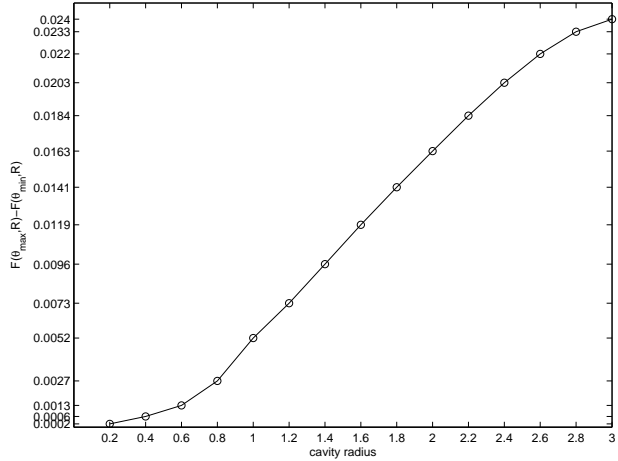}
\caption{Amplitude of oscillation.
$\mathcal{F}_c(90^\circ,R)-\mathcal{F}_c(0^\circ,R)$ }
\label{figure3}
\end{figure}

Our main results are shown in the two main curves of figure
\ref{figure3}. The figure \ref{figure2a} shows several curves of the
free energy density versus the angle $\theta$, each curve associated
to a different $R$ value. Similarly the figure \ref{figure2b} shows
several free energy density versus $R$ curves, each corresponding to
a distinct value of $\theta$. The figure \ref{figure2a} shows that
mirror symmetry $\theta \leftrightarrow \pi-\theta$ holds as
expected, since the two cavities inside the unit cell are
equivalent. The free energy $\mathcal{F}_c(\theta,R)$ is an even
function in $\theta$.

%

In our previous work\cite{DR04,RD04} we have found the remarkable
property that cavities inside a superconductor can lower its energy
as compared to the cavity-free superconductor. This effect can be
verified here in figure \ref{figure2b}, which display a set of
points lying under the energy threshold of $-0.434$, the free energy
$\mathcal{F}_c(\theta,R=0)$, of the system without cavities,
approximated by $\mathcal{F}_c(\theta=0,R=0.2)$, as previously
discussed. In summary we found here several pinning configurations,
each described by the pair of values $(\theta,R)$, that have lower
energy than the cavity free system. Table \ref{phasediagram}
exhibits these pairs $(\theta,R)$ of lower energy.
\begin{table}[h]
\caption{Pair of values $(\theta,R)$ that establishing a
configuration with lower energy than the cavity-free
superconductor.} \centering
\begin{tabular}{||c|c|c|c|c|c|c|c|c||}
 \hline
 $\theta$ (degree)& 0 - 3 &0 - 15&0 - 24&0 - 12&0 - 27&0 - 24&0 - 21&0 - 9\\
\hline
 $R$ ($\xi$) &0.4&0.6&0.8&1.0&1.2&1.4&1.6&1.8\\
\hline
\end{tabular}
\label{phasediagram}
\end{table}

The curve of figure \ref{figure2b} shows monotonic growth\cite{RD04}
for the free energy, but only for $R$ equal or larger than $1.2\xi$.
It also shows a local maximum for the free energy at $R=1.0\xi$.
In fact the behavior is distinct for the two regimes, $R>1.0\xi$ and $R\leq 1.0\xi$.
To understand the $R>1.0\xi$ case lut us hold the angle fixed, and vary the cavity radius.
One finds that the free energy density is given by a constand density
times the volume of the unit cell, $V_0=L^3$, removed of the
non-superconductor volume of the cavities, $V_c=8/3\pi R^3$.
This result is only approximately valid since the curvature of the order parameter near
the pinning sphere surface causes an increase in the kinetic energy,
an effect that becomes more pronounced for large spheres.
Thus we have that for $R>1.0\xi$,
\begin{eqnarray}
\mathcal{F}_c(\theta_{fixed},R)\propto\mathcal{F}_{cf}\left
(1-\frac{8\pi R^3}{3L^3}\right ), \label{eq:Rdependence}
\end{eqnarray}
where $\mathcal{F}_{cf}\cong\mathcal{F}_c(0^\circ,0.2\xi)=-0.434$ is
the cavity-free energy. The monotonic growth seen in figure
\ref{figure2b} is well described by the cubic $R$ dependence of
Eq.~\ref{eq:Rdependence}.
%
%

The dependence with the angle $\theta$ exhibited in figure
\ref{figure2a} may be described by the expression
\begin{eqnarray}
\mathcal{F}_c(\theta,R_{fixed})\propto f_k(R_{fixed})\sin^2{\theta}
\label{eq:thetadependece}
\end{eqnarray}
where the function $f_k(R)$ incorporate all the kinetics effects
produced by the presence of the cavities with radius $R$. In figure
\ref{figure2a}, $f_k(R)$ provides the amplitude of oscillation for a
fixed $R$. As we have mentioned before, the kinetic effects become
more pronounceable as the cavity radius increase.
We assume that $f_k$ is a linear function of $R$.

We add Equations \ref{eq:Rdependence} and \ref{eq:thetadependece}
together to obtain a general free energy expression to describe
arbitrary angles of rotation and radii bigger than the coherence length:
%
%
%
%
%
%
\begin{eqnarray}
\mathcal{F}_c(\theta,R)=\mathcal{F}_{cf}\left
(1-\frac{V_c}{V_0}\right )+f_k(R)\sin^2{\theta}. \label{eq:cavity}
\end{eqnarray}
For all radii, the configurations of minimum and maximum energies
are obtained for $0^\circ$ and $90^\circ$, respectively. The maximum
at $\theta=90^\circ$ has a smaller superconducting volume as
compared to the minimum at $\theta=0^\circ$ configuration. This is
easy to understand because the $0^\circ$ configuration has the two
spheres aligned along the z-axis and both overlap with the vortex
line, whereas the $90^\circ$ configuration only one overlaps the
vortex line. The other one is free in space thus taking away space
that could be otherwise superconducting. This makes the $90^\circ$
configuration closer to the normal state than the $0^\circ$ one.
At some intermediate angle between $0^\circ$ and
$90^\circ$ a depinning transition takes place, although it is not
noticeable in both figures \ref{figure2a} and \ref{figure2b}. This
transition has been studied in Ref.\cite{DR04}. The function
$f_k(R)$ is easily obtained by taking its difference at extreme
angles, $\mathcal{F}_c(\theta_{max},R)-\mathcal{F}_c(\theta_{min},R)$, where
$\theta_{max}=90^\circ$ and $\theta_{min}=0^\circ$. This difference
is shown in figure \ref{figure3} as a function of $R$.
Thus Eq.~\ref{eq:cavity}, the major conclusion of this paper,
gives a good description of the free energy for all $(\theta,R)$ pairs.
%

%

In the limit that $R\to 0$, the free energy should converge to the
cavity free superconductor, $\mathcal{F}_{cf}$, whose energy only
depends on the unit cell vorticity. Thus the term $f_k$ should
vanish in this limit $R\to 0$ so that the angular dependence
$\sin^2{\theta}$ disappears from the free energy:
\begin{eqnarray}
\lim_{R\to 0}f_k(R)&&\to 0. \label{eq:limits}
\end{eqnarray}
We determine the $f_k(R)$ term fitting the curve of the figure
\ref{figure3} with the best linear function. In this way, the
function found is:
\begin{eqnarray}
f_k(R)=-0.00358+0.00962 R \label{eq:F2}.
\end{eqnarray}
The linear dependence of the Eq.~\ref{eq:F2} with the radius
$R$ reflects the importance of the kinetic energy,
$\xi^2 \left|({\vec \nabla} -{{2\pi i}\over{\Phi_0}} {\vec A})\Delta \right|^2$,
which describes the bending of the order parameter
at the surface of the cavities, that becomes more important for large cavities.

Substituting all the terms in Eq. \ref{eq:cavity} by the terms
obtained in the fitting and the parameters of the system, the energy
dependence is expressed by following equation:
\begin{eqnarray}
\mathcal{F}_c(\theta,R)=\mathcal{F}_{cf}\left (1-\frac{8\pi
R^3}{3L^3}\right )+\left(0.00962 R - 0.00358\right)\sin^2{\theta}.
\label{eq:cavitynumerical}
\end{eqnarray}
The negative constant $-0.00358$ is in conflict with the condition
expressed by Eq.~\ref{eq:limits}. In fact in the region $R \leq
1.0\xi$, Eq.~\ref{eq:F2} does not apply because of mesh effects.
Besides for such small $R$ the present Ginzburg-Landau approach is
not applicable, as previously discussed.

\section{Many vortices near the zigzag of cavities}
\label{A general case}

\begin{figure*}[!t]
\centering
    \subfigure[]
    {
    \begin{minipage}[t]{0.46\textwidth}
    \centering
    \includegraphics[width=\linewidth]{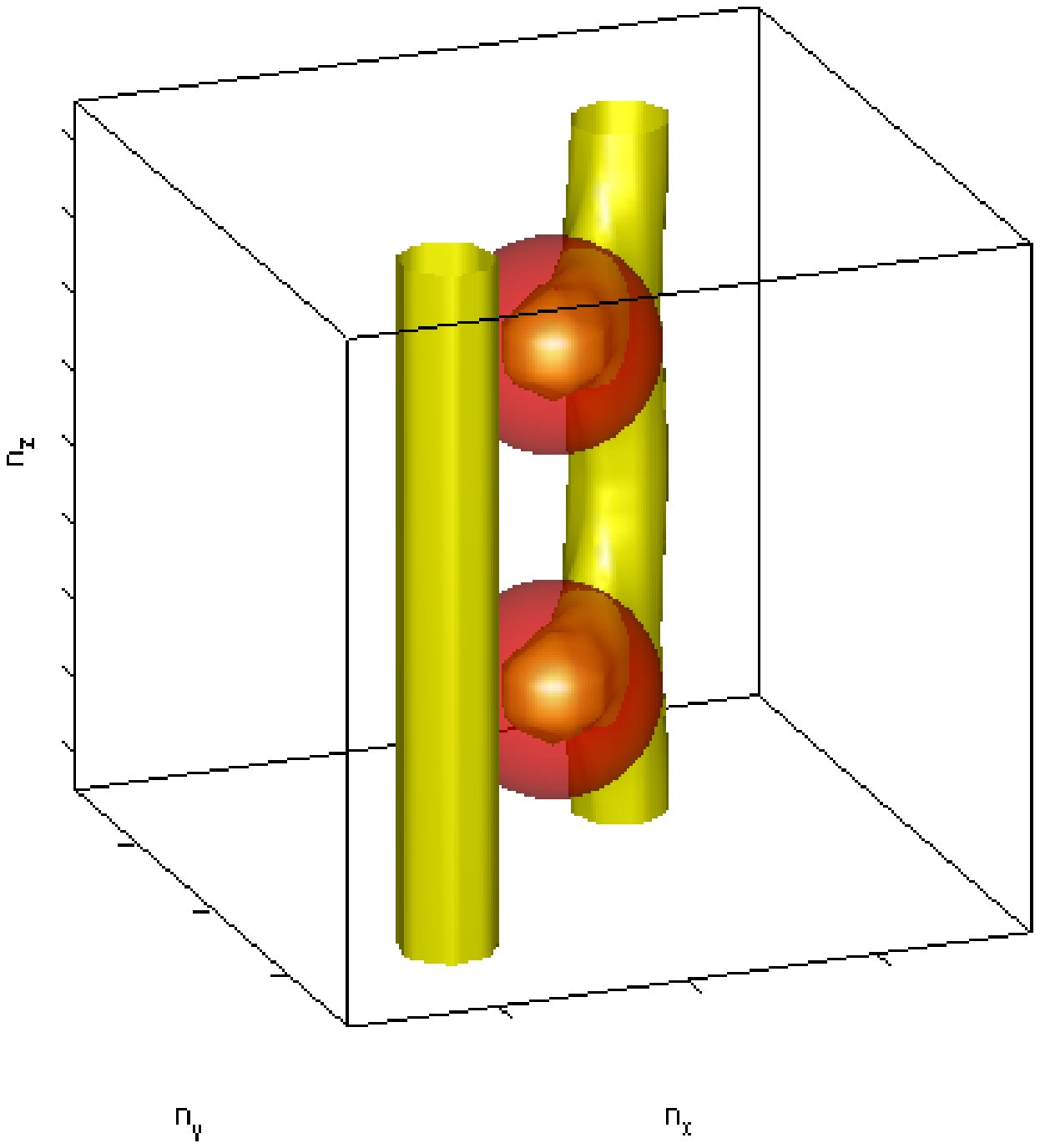}
    \label{figure4a}
    \end{minipage}
    }
    \subfigure[]
    {
    \begin{minipage}[t]{0.46\textwidth}
    \centering
    \includegraphics[width=\linewidth]{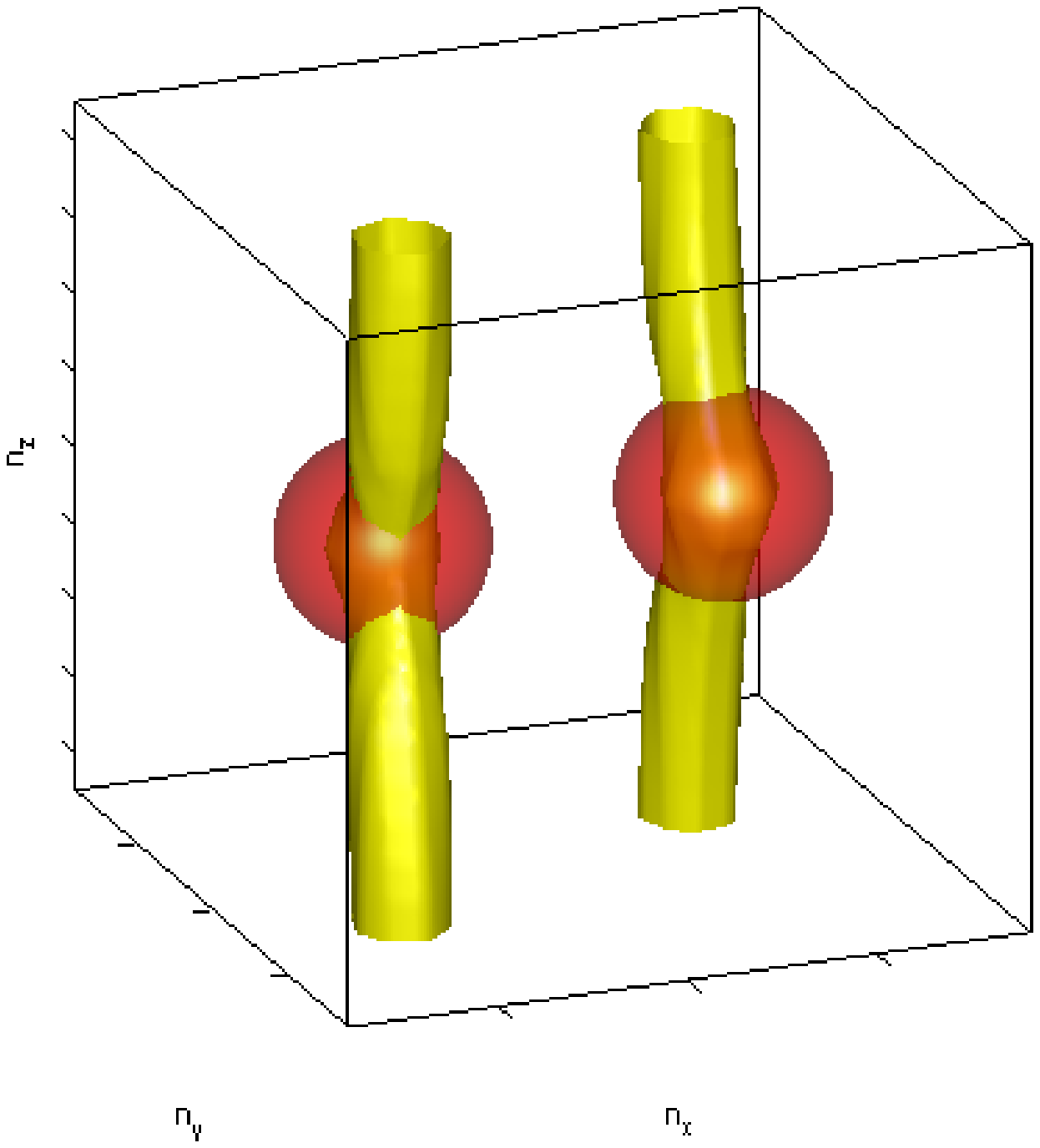}
    \label{figure4b}
    \end{minipage}
    }
    \caption
    {
    Iso-surfaces of the order parameter $|\Delta(\vec{x})|^2$
    inside the unit cell, a cube with side $L=12\xi$ containing cavities
    (in red) with radius $R=1.8\xi$. The figure \ref{figure4a}
    corresponds to $\theta=0^\circ$ and shows just one vortex line trapped by the defect,
    whereas the other one is not pinned and repelled due to vortex-vortex repulsion.
    In the figure \ref{figure4b} the cavities form an angle $\theta=90^\circ$ and
    each vortex is trapped by an independent cavity.
    }
    \label{figure4}
\end{figure*}

So far we have described the case of just one vortex near a zigzag
of cavities, $\vec{\nu}=1\hat{z}$. In this section we briefly
comment on the general case of many vortices along the z direction,
$\vec{\nu}=n_z\hat{z}$. As $n_z$ increases the upper critical field
is approached and beyond the upper critical field there is also a
surface superconductivity state at the surface of the cavities. The
radius of the cavities $R$ and the angle of rotation $\theta$ sets
new geometric configurations in the unit cell that lead to multiple
trapping and giant vortex states in the superconductor \cite{DZ02}.
A simple example of such configurations is shown in figure
\ref{figure4}, which displays two vortices in the unit cell ($n_z=2$
case) for $R=1.8\xi$ at the two extreme angles, $\theta=0^\circ$,
and $\theta=90^\circ$. For $\theta=0^\circ$ the system presents
strong competition between the vortex-vortex repulsion and the
vortex-cavity attraction. While one of the vortices is barely
trapped by the cavities the other vortex is not pinned. The rotation
of the cavities to the $\theta=90^\circ$ configuration yields a
state of two vortices each one trapped by a distinct cavity. For
values of $n_z$ bigger than 2 more complex situations are possible
and will be studied elsewhere.

\section{Conclusions}
\label{Conclusions}

We have carried here a Ginzburg-Landau theory study of a vortex line
near a zigzag of pinning centers. The pinning centers are insulating
spherical cavities and their arrangement is well described by a unit
cell containing two of them. Inside the unit cell the rotation of
the two cavities around the center produces a continuous of zigzag
arrangements whose effects on the vortex line are here discussed. We
observe distinct behavior for cavity radius above and below
$1.0\xi$. Sweeping the angle of rotation of the cavities around its
center, $\theta$, makes the zigzag more pronounced to the point that
the vortex line decouples from it above the critical angle
$\theta_c$, as previously found\cite{DZ02}. Below this angle the
vortex line is pinned by both cavities and above by just one. Here
we have determined that the free energy density has a simple
analytical dependence, expressed by the Eq.~\ref{eq:cavity}.
\section{Acknowledgments}
\label{Acknowledgments}

Research supported in part by Instituto do Mil\^enio de
Nano-Ci\^encias, CNPq, and FAPERJ (Brazil).



\bibliographystyle{apsrev}
\bibliography{DRBJP07}
-----------------------------
\end{document}